\documentstyle[sprocl]{article}

\def\be{\begin{equation}}
\def\ee{\end{equation}}
\def\bea{\begin{eqnarray}}
\def\eea{\end{eqnarray}}

\begin{document}

\begin{titlepage}
\title{HARD THERMAL LOOPS IN THE ELECTROWEAK THEORY}
\vspace{1cm}

\author{ CRISTINA MANUEL \\
CERN TH-Division\\
CH-1211 Geneva 23, Switzerland}

\maketitle
\vspace{1.cm}
\def\baselinestretch{1.15}
\begin{abstract}
We derive a thermal effective action for soft fields in the 
broken phase of the electroweak theory in the limit of a strongly interacting 
Higgs sector. This action is just the proper generalization of the hard thermal
loop effective action of a Yang-Mills theory when there is a Higgs mechanism and for
a heavy Higgs boson. One can obtain from this action the thermal corrections to
the masses of the $W$, $Z$ and the photon.
\end{abstract}
\vspace{4cm}
\leftline{CERN-TH/99-42}
\leftline{February 1999}

\thispagestyle{empty}

\vskip-15.cm
\rightline{{ CERN-TH/99-42}}
\rightline{{ hep-ph/9902444}}

\vskip3in

\end{titlepage}

\title{HARD THERMAL LOOPS IN THE ELECTROWEAK THEORY}

\author{CRISTINA MANUEL}

\address{Theory Division, CERN\\
CH-1211, Geneva 23, Switzerland\\ 
E-mail: Cristina.Manuel@cern.ch}


\maketitle\abstracts{We derive a thermal effective action for soft fields in the 
broken phase of the electroweak theory in the limit of a strongly interacting 
Higgs sector. This action is just the proper generalization of the hard thermal
loop effective action of a Yang-Mills theory when there is a Higgs mechanism and for
a heavy Higgs boson. One can obtain from this action the thermal corrections to
the masses of the $W$, $Z$ and the photon.}

In a non-Abelian plasma, such as the one formed 
in the high temperature $T$ regime of QCD or in the
restored phase of the electroweak theory, 
the effective theory for the soft modes
is given by a sum of the tree level action plus the
the hard thermal loop one \cite{BP}. Soft refers to a physical scale of order $\sim g  T$,
where $g$ is the gauge coupling constant.
The hard thermal loop (HTL) effective action $S_{HTL}$
 is understood as arising
after integrating out the  modes of order $\sim T$ of the theory.
The effective action $S_{HTL}$ gives account of the by now well 
understood phenomena of Debye screening and Landau damping.
It has been derived using different approaches in the literature.
The first and fundamental one was based on an evaluation of individual 
Feynman diagrams \cite{BP}, and it was essential for all the subsequent developments,
as it allowed to recognize the importance of  the hierarchy of scales $T$ and $gT$ in
the plasma.
After realizing that all the HTL's were related by simple Ward identities \cite{Ward},
it was then discovered the underlying gauge symmetry principle of the 
infinite set of HTL's \cite{gaugeinv}. The fact that $S_{HTL}$ is a gauge invariant action
was exploited in a subsequent set of papers \cite{others}.

A different language to describe HTL's, that of kinetic theory, was then introduced soon after.
Blaizot and Iancu \cite{BI} derived a set of kinetic equations from the hierarchy of Schwinger-Dyson
equations obtained from the QCD Lagrangian. There is a different approach to yield
the same kinetic equations which has as a starting point the Wong equations \cite{Wong}. The Wong
equations are the classical equations of motion of point particles carrying
a non-Abelian charge. A transport theory can be constructed from those Wong
equations \cite{Heinz}, and in the response theory, again it reproduces the HTL's \cite{KLLM}.

We now understand  the physics at soft scales in the symmetric phase of a hot
non-Abelian gauge field
theory, and we also have different languages to describe the
same thermal physical phenomena. The effective action $S_{HTL}$ is essentially a mass
term for the chromoelectric fields. Chromomagnetic fields are not screened at 
the same order in the coupling constant, although there is the strong belief that 
a mass for the magnetic degrees of freedom is generated non perturbatively.

In this talk we will derive an effective action
for soft fields in a theory where the non-Abelian gauge fields are
already massive at $T=0$. 
We will then concentrate our attention to the electroweak theory in its broken phase.
And more particularly, we will consider the limit where the Higgs field becomes heavy, and then
the Higgs sector becomes strongly interacting. If the Higgs boson is very heavy, one
can integrate out the Higgs field from the electroweak Lagrangian to finally obtain
a low energy effective theory which reads \cite{Appel}
\be
\label{eq1} 
{\cal L}_{eff}  = - \frac 12 \,{\rm Tr} (W_{\mu \nu} W^{\mu \nu} ) - 
\frac 14\, B_{\mu \nu} B^{\mu \nu} +
 \frac {v^2}{4} {\rm Tr} \, (D_\mu \Sigma\, D^\mu \Sigma ^\dagger)\ .
\ee
(see Ref. \cite{cm1}  for notations and conventions).

This low energy effective theory is interesting due to several reasons.
The experimental lower bound for the Higgs mass is continuously increasing ( now is above 90 GeV),
so the possibility of a heavy Higgs boson is not yet excluded. And even if the Higgs
boson does not exist, the above Lagrangian would also describe the low energy theory
of no matter what mechanism is responsible for the generation of masses of the 
gauge bosons. At tree level the model is universal, and thus model independent.
The model dependence only arises at the one-loop level. For example, if one wants to
include the effects of quantum Higgs boson exchange, additional effective interactions
are required, thus \cite{Appel}
\be
\label{eq2} 
{\cal L}_{4}  = L_1 {\rm Tr} \, (D_\mu \Sigma\, D^\mu \Sigma ^\dagger)^2 + 
L_2 {\rm Tr} \, (D_\mu \Sigma\, D^\nu \Sigma ^\dagger)^2 \ ,
\ee
where $L_{1,2}$ are model dependent constants.  The above terms are
higher order in an energy expansion. 

We will only consider the thermal effects associated to the lowest (universal)
effective Lagrangian Eq. (\ref{eq1}) after integrating out the hard modes of the theory.
We will not consider the effect of fermions for the time being.
   
In the Lagragian Eq. (\ref{eq1}) there are three types of fields: the weak isospin field
$W_\mu$, the weak hypercharge field $B_\mu$ and the would-be Goldstone bosons $\phi^a$ which
are contained in the matrix $\Sigma$, and 
which are unphysical. This is apparent in the unitary gauge $\Sigma =1$, where the
Goldstone fields disappear. In this gauge one can easily read  the masses
of gauge bosons  $W$, the $Z$ and the photon,  
to reach to the  well-known values 
\be 
\label{eq3}
M_W = \frac{g v}{2}, \qquad M_Z = \frac{\sqrt{g^2 + g'^2} v}{2} \ , \qquad M_{\gamma} = 0 \ .
\ee

For loop computations it is however much more convenient to work in a different gauge,
since in the unitary gauge the gauge field propagators are not well behaved in the 
infrared region.

To integrate out the hard modes of the theory it is convenient to use the background
field method (BFM) \cite{BFM}. The BFM is a well-known technique to evaluate loop effects, and
consists in expanding the tree level Lagrangian around the solution of the classical
equations of motion. The one-loop effective action is then obtained after integrating out
the quantum fluctuations. The BFM has certain advantages when applied to gauge field 
theories, as we will see. After the vector gauge fields are split into background
(or classical) and quantum pieces, there are two types of symmetries associated to the tree level
action. One is associated to the background fields, and the other one is associated
to the quantum fields. One can then fix the gauge of  the two different symmetries 
independently, as long as the quantum gauge is fixed in a covariant way with respect
to the  background gauge symmetry. The one-loop effective action thus obtained  is
automatically respectful with the background gauge symmetry \cite{BFM}.

In the spirit of the BFM, one then introduces an additive splitting for the
gauge fields, and a multiplicatively one for the matrix $\Sigma$ \cite{cm1}.
At that point, one can perform a Stueckelberg transformation \cite{Stu}
to get rid of the  the background Goldstone fields, which is equivalent to choosing the unitary
background gauge, while choosing a different gauge for the quantum fields.  
In the unitary background gauge one then finds \cite{cm1}
\begin{eqnarray}
\label{4.1}
S_{eff} & + & \delta S_{eff,T}  =   \\
& = &\int d^4 x \left\{
- \frac 12 \,{\rm Tr} ({\bar W}_{\mu \nu} {\bar W}^{\mu \nu} ) - 
\frac 14 \, {\bar B}_{\mu \nu} {\bar B}^{\mu \nu} + \frac {v^2(T)}{4} 
\,{\rm Tr}\left( g {\bar W}^\mu - g' {\bar B}^\mu \frac{\tau^3}{2} \right)^2 \right\}
\nonumber \\
&- &\frac{T^2}{6}\int \frac{d \Omega_{\bf q}}{4 \pi} \int d^4 x \,d^4 y\,
{\rm Tr} \left({\bar \Gamma}_{\mu \lambda} (x) <x | \frac{Q^\mu Q_{\nu}}{- (Q \cdot
{\bar d})^2} |y> {\bar \Gamma}^{\nu \lambda} (y) \right) \nonumber \\
& + &  \frac{g^2 T^2}{3}\int \frac{d \Omega_{\bf q}}{4 \pi} \int d^4 x \,d^4 y\,
{\rm Tr} \left({\bar W}_{\mu \lambda} (x) <x | \frac{Q^\mu Q_\nu}{- (Q \cdot
{\bar D}_W)^2} |y> {\bar W}^{\nu \lambda} (y) \right)
\nonumber
\end{eqnarray}
where ${\bar W}_{\mu \nu}$,  ${\bar B}_{\mu \nu}$ are the field strengths
of the corresponding background gauge fields, and
\begin{eqnarray}
\label{4.2}
v(T) & =  & v \left( 1 - \frac{1}{12} \frac {T^2}{v^2} \right) \ , \\
\label{4.3}
{\bar \Gamma}_\mu & = & \frac i2 \left( g {\bar W}_\mu + g' {\bar B}_\mu \frac{\tau^3}{2}\right)\ , \\
{\bar \Gamma}_{\mu \nu} & = & \partial_\mu {\bar \Gamma}_\nu
- \partial_\nu {\bar \Gamma}_\mu +
 [{\bar \Gamma}_\mu,{\bar \Gamma}_\nu] \ , 
\end{eqnarray}
and $Q= (i, {\bf q})$ is a null vector $Q^2=0$. The angular
integral in (\ref{4.1}) is done over all directions of the three 
dimensional unit vector ${\bf q}$. One can obtain the same action in an
arbitrary background gauge just by inverting the Stueckelberg transformation \cite{cm1}.

The above effective action is the proper generalization of the 
HTL effective action in the broken phase of the electroweak theory in the limit
of a heavy Higgs field. 
One can obtain the thermal screening masses for all the gauge fields by considering the
above action in the static limit. One then finds the following values for the transverse
masses
\begin{equation}
\label{4.x}
M^2_{W,t} (T) = \frac{g^2 v^2 (T)}{4} \ , \qquad M^2_{Z,t}(T) = \frac{(g^2 + g'^2) v^2 (T)}{4} \ , 
\qquad  M^2_{\gamma, t} (T) = 0   \ ,
\end{equation}
while  for the longitudinal masses one has
\begin{eqnarray}
\label{4.xx}
M^2_{W,l} (T) & = & \frac{g^2 v^2 (T)}{4} + \frac{3 g^2 T^2}{4}
\ , \qquad M^2_{\gamma, l} (T)  =  e^2 T^2 \nonumber \\
 M^2_{Z,l}(T) & = & \frac{(g^2 + g'^2) v^2 (T)}{4} + 
\frac{  (g^2+g'^2) T^2}{12} \left( \cos^2{\theta_W} -\sin^2{\theta_W}\right)^2  \nonumber \\
& + & \frac{2  g^2 T^2}{3} \cos^2{\theta_W}  \ ,
\end{eqnarray}
where $\theta_W$ is the Weinberg angle.
The above results agree with those computed in Ref.
\cite{Gavela}. Notice that the transverse and longitudinal modes get different thermal
corrections to their masses.

Let us comment the temperature range of validity of the previous results. The above
effective action has been derived assuming that $M_W \ll T \ll \sqrt{12}v$. The lower
limit just tells us that the masses of the gauge bosons are soft.
The upper limit is
just demanded in order to require that the one-loop effects computed here are small 
corrections. If this is not the case, then one cannot guarantee that the next to leading
order effects are not important. The next to leading order effects would arise by
considering two loops from Eq. (\ref{eq1}), and the (model dependent) one loop effects generated
by Eq. (\ref{eq2}).

Let us finally conclude with the observation that it would be very useful
to derive kinetic equations which describe these thermal effects in the broken
phase of the electroweak theory.

\section*{Acknowledgments}
I would like to thank the organizers of this conference for their hospitality and for creating a nice
atmosphere for very interesting physical discussions.

\end{document}